\documentstyle[aps]{revtex}

\input boxedeps
\HideDisplacementBoxes

\SetRokickiEPSFSpecial          

\begin{document}
\draft
\twocolumn[\hsize\textwidth\columnwidth\hsize\csname @twocolumnfalse\endcsname


\title{Internal Structure of Einstein--Yang--Mills Black Holes}
\author{E. E. Donets}
\address{Laboratory of High Energies, JINR, 141980, Dubna, Russia
}
\author{D. V. Gal'tsov}
\address{Department of Theoretical Physics,
 Moscow State University, 119899, Moscow, Russia
}
\author{M. Yu. Zotov}
\address{Nuclear Physics Institute, Moscow State University,
 Moscow 119899, Russia
}
\date{\today}
\maketitle

\begin{abstract}
 It is shown that a generic black hole solution of the $SU(2)$
 Einstein--Yang--Mills (EYM) equations develops a new type of an
 infinitely oscillating behavior near the singularity. Only
 for certain discrete values of the event horizon radius exceptional
 solutions exist possessing an inner structure of
 the Schwarzschild or  Reissner--Nordstr\"om  type.

\end{abstract}
]
\narrowtext

 Discovered soon after the regular Bartnik--McKinnon (BK) solutions \cite{bk},
 EYM black holes (BH) \cite{vgkb,vg,bfm} provided new insights into
 the BH physics related to the no--hair and uniqueness theorems
 \cite{todos}. They share sphaleronic properties of the BK particle--like
 solutions \cite{sph} and also exhibit an unusual discreteness
 (`quantization' of the YM field on the event horizon) due
 to a singular non--linear boundary value problem in the domain
 between the horizon and the asymptotically flat (AF) infinity. Still,
 the existing knowledge of the EYM BH's (unlike the BK objects)
 is incomplete since only
 external solutions have been constructed so far (though a qualitative
 discussion of inner solutions is available \cite{vg}).
 Here we present brief results of our investigation of the interior
 structure of the EYM BH's which reveal new surprising features due
 to coupling of non--linear fields to gravity.

 Assume the static spherically symmetric magnetic ansatz for the
 YM potential
\[
 A=\left(W(r)-1\right) \left(T_{\varphi}d\theta-
 T_{\theta}\sin\theta d\varphi\right),
\]
 ($T_{\varphi, \theta}$ are
 spherical projections of the $SU(2)$ generators)
 and the following parametrization of the metric
\begin{equation}
\label{ds2}
 ds^2 = (\Delta/r^2) \sigma^2dt^2-
 (r^2/\Delta) dr^2-r^2d\Omega^2,
\end{equation}
 where $d\Omega^2 = d\theta^2 + \sin^2\theta d\varphi^2$, and
 $\Delta$, $\sigma$ depend on $r$.

 The field equations include a coupled system for $W$, $\Delta$
\begin{eqnarray}
 & \Delta (W'/r)' + F W' = W V/r, & \label{eq1}  \\
 & (\Delta/r)' +  2 \Delta (W'/r)^2 =F ,&\label{eq2}
\end{eqnarray}
 where $V=(W^2-1)$, $F=1 - V^2/r^2$, and a decoupled equation for $\sigma$:
\begin{equation} \label{eq3}
 (\ln\sigma)' = 2W'^2/r.
\end{equation}

 These equations admit BH solutions in the domain
 $r\ge r_h$ for any  radius of the event horizon $r_h$. The solutions
 are specified by the number $n\in N$ of nodes of $W$ thus forming
 a discrete set for each $r_h$. Although it is not guaranteed
 {\it a priori} that the chart (\ref{ds2}) is extendible
 to the full region $r<r_h$, for AF
 solutions we did not meet any singularity in the interior region
 unless the genuine one $r=0$ is reached. In terms of coordinates
 (\ref{ds2}) one can find three distinct local power series solutions.
 The first one is Schwarzschild--like (S), it corresponds to the
 vacuum value of the YM field $|W(0)|=1$. Using the mass function
 $m(r),\,\Delta=r^2-2mr$, one gets \cite{vg}
\begin{eqnarray} \label{s}
 &&W=-1+b r^2 +b^2(3-8b)r^5/(30m_0) + O(r^6),\nonumber\\
 &&m=m_0 (1-4b^2r^2+8b^4r^4)+2b^2r^3+O(r^5),
\end{eqnarray}
 where $m_0$, $b$ are (the only) free parameters.

 The second is the Reissner--Nordstr\"om (RN) type of solution which can be
 found assuming the leading term of $\Delta$ to be a positive constant.
 This  requires $W(0)= W_0\ne \pm 1, 0$ and gives \cite{vg}
\begin{eqnarray}\label{rn}
 &&W=W_0-W_0 r^2/(2V_0)+c r^3/(2W_0V_0)+O(r^4),\nonumber\\
 &&\Delta=V_0^2  - 2 m_0r + 2(c+m_0W_0^2/V_0^2)r^3+O(r^4),
\end{eqnarray}
 what corresponds to the RN metric of the mass $m_0$ and the (magnetic)
 charge $P^2=V_0^2$, $V_0=V(W_0)$. The expansion contains three free
 parameters $W_0$, $m_0$, $c$.

 We have also found the third local power series solution assuming a
 {\it negative} value for $\Delta(0)$ (i.e. {\it imaginary} $P$):
\begin{eqnarray} \label{new}
 &&W=W_0 \pm r - W_0 r^2/(2V_0) +O(r^3), \nonumber\\
 &&\Delta=-V_0^2 \mp 8W_0r/V_0+O(r^2),\\
 &&\sigma=\sigma_1 (r^2\mp 4W_0 r^3/V_0) + O(r^3).\nonumber
\end{eqnarray}
 Here there is only one free parameter $(W_0)$ for $W$, $\Delta$.
 The corresponding space--time near the singularity
 is conformal to the cylinder (after a time rescaling):
\begin{equation}
 ds^2=r^2(dt^2 -dr^2 - d\theta^2-\sin^2\theta d\varphi^2).
\end{equation}

 However, one may suspect that such asymptotics can not correspond to
 a {\it generic} BH. Imposing `boundary conditions' in the singularity  we
 obtain the same kind of the singular boundary value problem as one for
 external solutions where a similar role is played by the AF
 condition. The latter is known to result in the `quantization' of the allowed
 values $W^n(r_h)$. Internal 'boundary value' problem leads to the second
 `quantization', now of the event horizon radius $r_h$.
 Therefore, EYM BH's with S and RN interiors may constitute only the set
 of zero measure in the whole EYM BH solution space. For a generic $r_h$
 the interior metric is strikingly different.

\begin{figure}[htbp]
\begin{center}
 {\BoxedEPSF{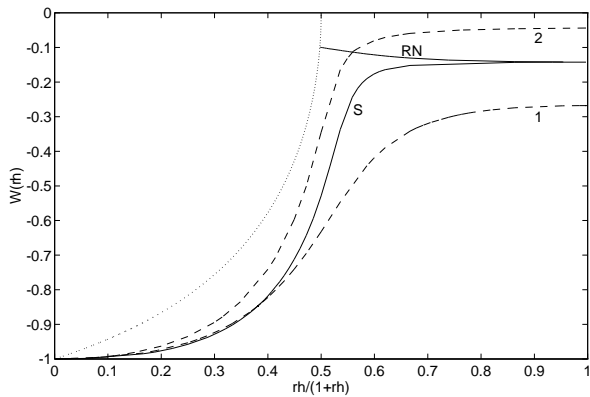 scaled 465}}
\end{center}
\caption{
 $W(r_h)$  for the S-- and RN--type interior solutions. Dashed lines --
 $W^n(r_h)$ for $n=1, 2$ (higher--$n$ curves lie
 between  the $n=2$ one and the boundary $r_h=1-{W_h}^2$, dotted line).
 Note that S and RN curves $W(r_h)$ do not merge.
 }
\label{alaBFM}
\end{figure}

\begin{figure}[htbp]
\begin{center}
 {\BoxedEPSF{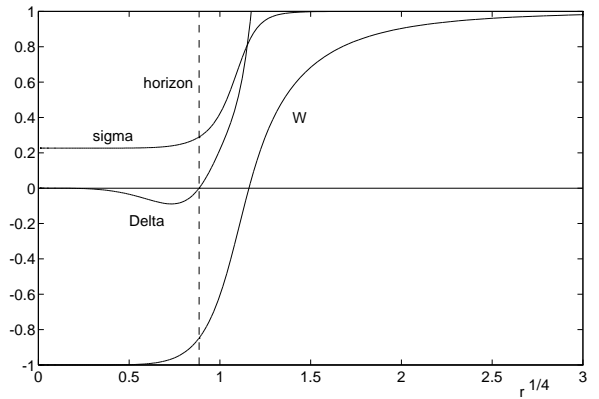 scaled 475}}
\end{center}
\caption{The $n=1$ EYM BH (S--type).}
\label{Schw}
\end{figure}

 The system (\ref{eq1}--\ref{eq2}) was integrated numerically
 in the region $0<r<r_h$ using an
 adaptive step size Runge--Kutta method for various $r_h=10^{-8},...,10^{6}$
 starting at the left vicinity of the event horizon $r_h$ with one
 free parameter $W_h = W(r_h)$ satisfying inequalities  $|W_h|<1$ and
 $1-W_h^2<r_h$ which are the necessary conditions for asymptotic flatness.
 For given $r_h$, the interior solutions meeting the
 expansions (\ref{s}--\ref{new}) may exist only for
 some appropriate $W_h$. A numerical strategy used to find
 such $W(r_h)$ consisted in detecting the change of sign of the derivative
 $W'$. In the S-case we found the curve $W(r_h)$ which
 starts at $-1$ as $r_h\rightarrow 0$ and approaches $-0.1424125$ for large
 $r_h$ (Fig.\ \ref{alaBFM}) (without loss of generality we choose
 $W_h<0 $).  Our S-curve intersects the
 $n=1$ member of the family $W^n(r_h)$, corresponding to the set
 of external AF solutions.  Parameters of this BH solution are
 shown in Tab.\ \ref{numres}, its global behavior is depicted
 in Fig.\ \ref{Schw}.
\footnote{In later publication by Breitenlohner
 et al. (gr-qc/9703047) S-solution with $n=3$ was found, so
 our S-solution is not unique as was claimed in the first version 
 of this paper. We also eliminate here an 
 irrelevant  for BH solutions 'microstructure' of the 
 RN curve near the boundary (shown previously in Figs. 1,4)
 which was  perhaps a numerical artefact.}

\vspace{7ex}
\vspace{4ex}
\begin{figure}[htbp]
\begin{center}
 {\BoxedEPSF{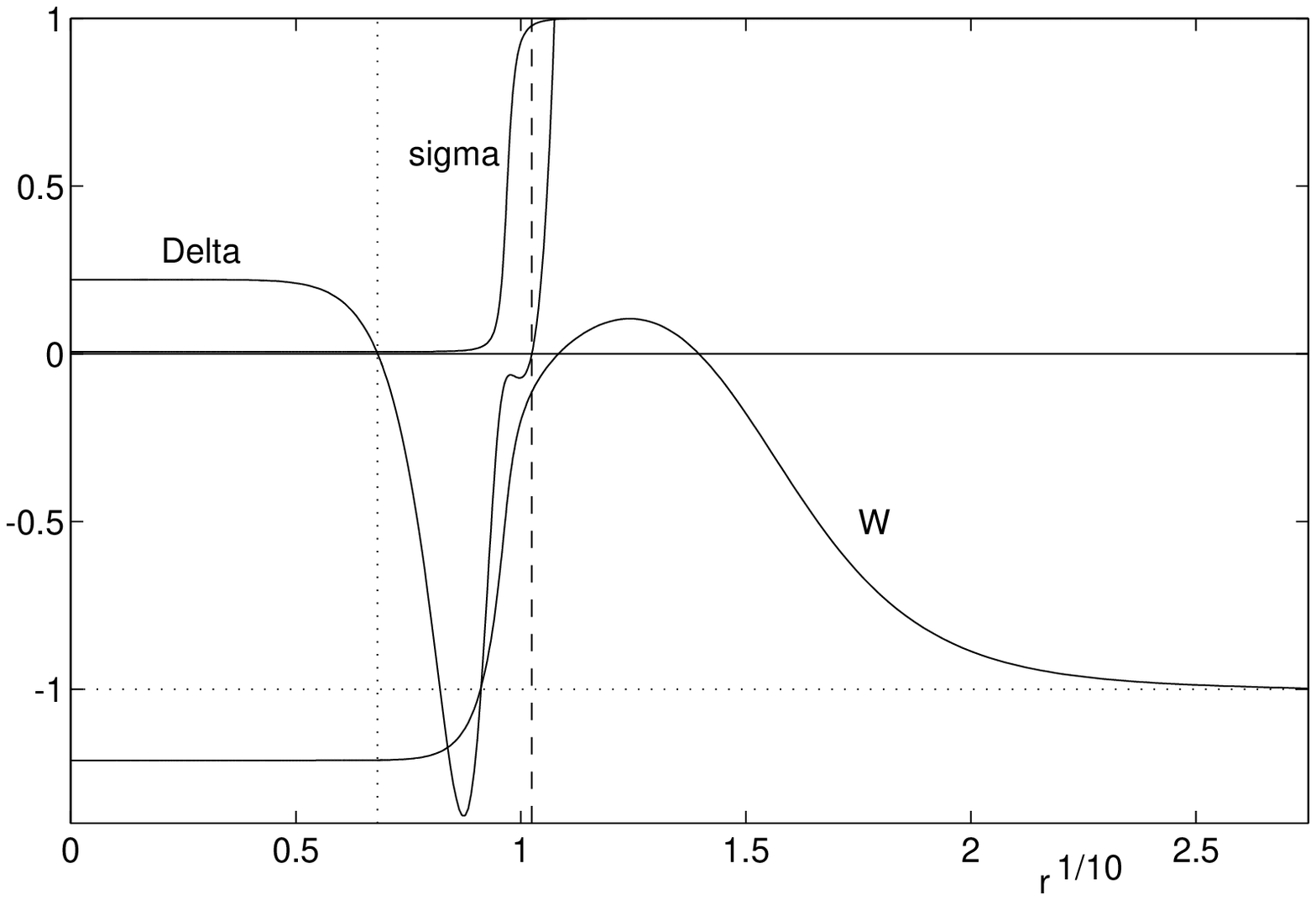 scaled 480}}
\end{center}
\caption{The $n = 2$ EYM BH (RN--type).}
\label{RN2}
\end{figure}
\vspace{4ex}
\begin{figure}[htbp]
\begin{center}
 {\BoxedEPSF{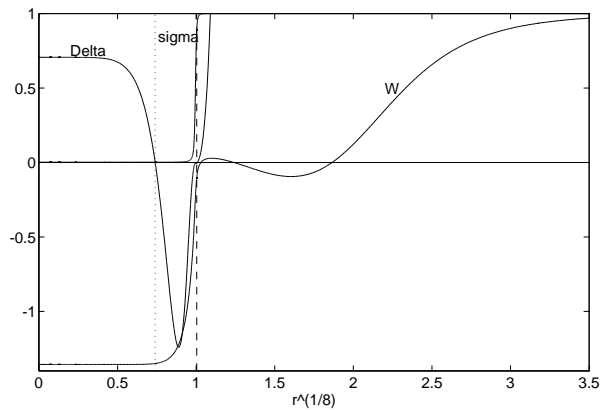 scaled 480}}
\end{center}
\caption{The $n = 3$ EYM BH (RN--type).}
\label{RN3}
\end{figure}

 Interior solutions of the RN--type, meeting the expansions
 (\ref{rn}) in the singularity, were found for $r_h >r_h^*=0.990288617$.
 The corresponding curve $W(r_h)$ (also shown in Fig.\ \ref{alaBFM})
 intersects the curves $W^n(r_h)$ for all $n\ge 2$.
 These solutions possess an inner Cauchy horizon at some $r_{\_} < r_h$ with
 $|W(r_{\_})|>1$ (Figs.\ \ref{RN2}, \ref{RN3}).

\begin{figure}[htbp]
\begin{center}
 {\BoxedEPSF{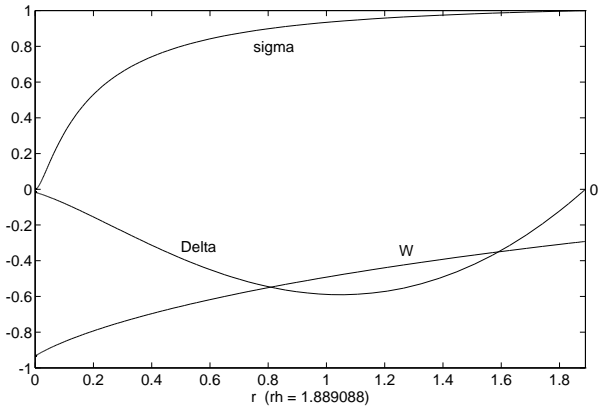 scaled 480}}
\end{center}
\caption{Interior solution meeting expansions (7) at $r=0$.}
\label{unique}
\end{figure}

 Solutions of the third type (\ref{new}) were studied numerically
 starting from the vicinity of the origin.
 The unique solution has been found for the horizon data subject to
 conditions $|W_h|<1$ and $1-W_h^2<r_h$ for the upper sign in (\ref{new}) and
 $W(0)=-0.9330656$, corresponding to $r_h=1.889088$
 (Fig.\ \ref{unique}). This solution, however, does not meet any value
 $W^n(r_h)$ and thus does not present a BH.

\vspace{6ex}

\begin{figure}[htbp]
\begin{center}
 {\BoxedEPSF{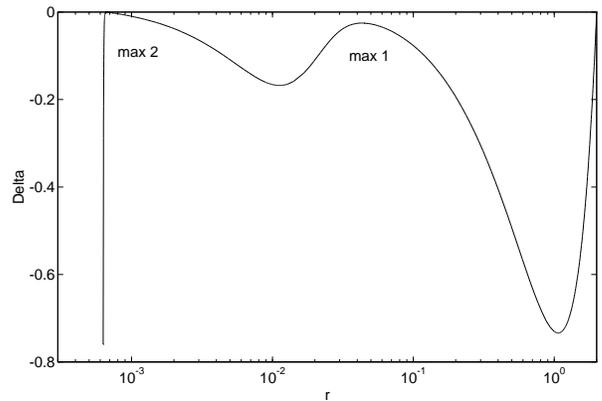 scaled 470}}
\end{center}
\caption{The beginning of $\Delta$--oscillations  for $n=1$,
 $r_h=2$, $W(r_h)=-0.342072$.}
\label{fall}
\end{figure}

 Hence, we have found that the EYM BH's with the `standard' interiors
 of the S and RN types exist only for certain discrete values of $r_h$.
 For all other (continuously varying) $r_h$ one observes oscillations of
 $\Delta$ in the lower half--plane with an
 infinitely growing amplitude near the singularity.
 The oscillation region starts with an exponential fall of $\Delta$
 which typically occurs after passing a local maximum $r^{max}$
 (Fig.\ \ref{fall}).
 In this regime the right hand side of
 (\ref{eq1}) becomes comparatively small with respect to other
 terms, and one can simplify the system analytically. Neglecting the
 right side of (\ref{eq1}), one obtains the following
 (approximate) first integral of the system (\ref{eq1}--\ref{eq3})
\begin{equation}
\label{z}
 Z=\Delta W' \sigma/r^2=const.
\end{equation}
 In what follows we will describe in more detail the lower--$n$ case
 (qualitatively the behavior of $\Delta$ in the oscillating regime is
 the same for all $n$). Starting from  some point
 $r_0$ close to $r^{max}\,(r_0\lesssim r^{max}$) the quantity $U=W'/r$
 becomes approximately constant $U=U_0\gg 1/r^{max}\gg1$
 ($r^{max}\ll 1$ for lower $n$).
 Then, using the Eqs.\ (\ref{eq2}) and (\ref{z}), one finds the following
 estimate for $\Delta$ valid in the region of the exponential
 fall \cite{mz}
\begin{equation}\label{de}
 \Delta=-Cr \exp\left[-(U_0 r)^2\right],
\end{equation}
 where $C$ is a positive constant. From this expression it is clear that
 the fall must stop at the local minimum
\[
 r^{min}=(\sqrt{2}\;|U_0|)^{-1}.
\]

\vspace{3ex}
\begin{figure}[htbp]
\begin{center}
 {\BoxedEPSF{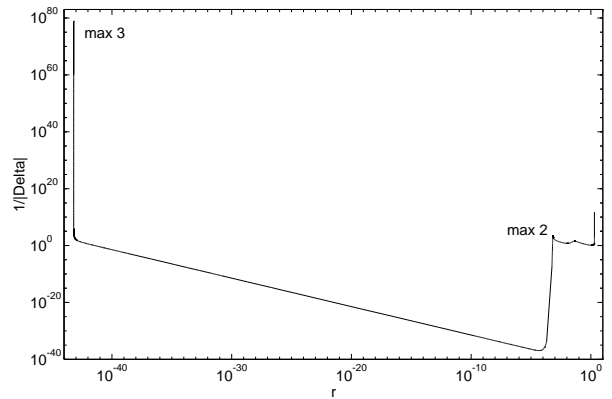 scaled 470}}
\end{center}
\caption{  $|\Delta|^{-1}$ for the second oscillation cycle of
 $n=1$ BH solution with $r_h=2$, $W_h=-0.342072$. }
\label{Delta}
\end{figure}

 According to (\ref{z}, \ref{de}), $\sigma$ decreases exponentially during
 the fall of $\Delta$:
\[
 \sigma^{min}\sim \sigma^{max} \exp\left[-(U_0 r^{max})^2\right],
 \quad U_0 r^{max}\gg 1,
\]
 (typically $r^{min}\ll r^{max}$). After  $r^{min}$ is passed,
 $U$ still remains almost unchanged, and hence the exponential factor
 in (\ref{de}) becomes irrelevant. It follows that after passing
 the minimum, $\Delta$ starts
 to grow linearly so that $|\Delta/r|\approx 2m$ is constant. This regime
 breaks down when $|\Delta|$ becomes comparable with $V^2$. Then a
 rapid growth of $U$ takes place, while $\sigma$ is still changing slowly.
 At this stage, according to (\ref{z}), the product $U\Delta$ remains
 almost constant. Finally
 $\Delta$ reaches the next local maximum (asymptotically points of
 local maxima approach zero), and then a new oscillation
 period starts with a grown up amplitude (Tab.\ \ref{numosc},
 Fig.\ \ref{Delta}). It is clear from (\ref{eq3}) that $\sigma$ monotonically
 decreases exhibiting rapid falls during exponential
 falls of $\Delta$ and keeping almost constant values while $\Delta$ is
 growing up. Thus, $\sigma$ tends to (but does not reach) a zero limit
 as $r\rightarrow 0$.

\begin{figure}[htbp]
\begin{center}
 {\BoxedEPSF{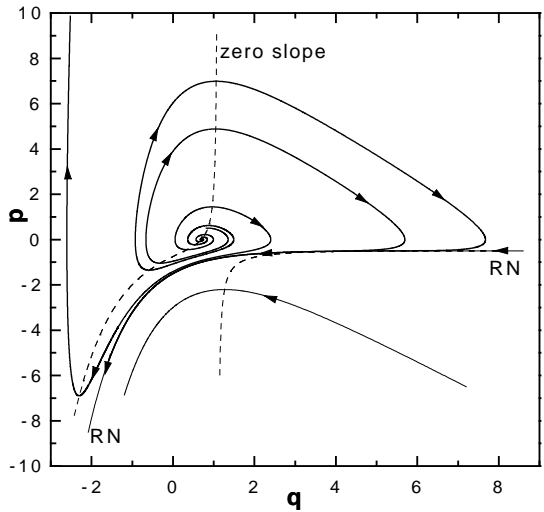 scaled 420}}
\end{center}
\caption{Phase portrait of (11), {\sf RN}~-- an invariant set
 corresponding to the RN--type solution (dashed -- zero slope lines).}
\label{main}
\end{figure}

 Although the derivative $W'$ takes
 rather large absolute values on some intervals, the corresponding
 variation of $W$ is still small because these
 intervals are also extremely small.
 All the above features (small variation of $Z$ and $W$, constancy of $U$
 while $|\Delta|$ is falling down) become more pronounced while
 oscillations progress implying that both $Z$ and $W$ have
 finite limits as $r\rightarrow 0$.
 Then neglecting the right side of (\ref{eq1}),
%
%
 omitting $1$ in $F$, and replacing $W$ by its limiting value $W_0$,
 one arrives at the following two--dimensional
 autonomous dynamical system
\begin{eqnarray}
 &&{\dot q}=p,\nonumber\\
 &&{\dot p}=(3e^{-q} -1)p+2e^{-2q}-1/2,
\end{eqnarray}
 where $\Delta=-(V_0^2/2) \exp(q)$,
%
%
 and a dot stands for derivatives with respect to $\tau=2 \ln (r_h/r)$.
 This system has one (focal) fixed
 point ($p=0$, $q=\ln 2$) with eigenvalues $\lambda=(1\pm i\sqrt{15})/4$, its
 phase portrait is shown in Fig.\ \ref{main} together with an invariant set
 $p = - e^{-q} - 1/2$ corresponding to the RN--type solution. The oscillating
 solutions lie above this curve.
 The phase motion in this region is unbounded, and there are
 no limiting circles. The limit $q=-\infty$ ($\Delta=0$) can not
 be reached, $\Delta$ remains negative valued
 as $r \rightarrow 0$ and passes an infinite sequence of local maxima
 and minima (Fig.\ \ref{main}).

\vspace{4ex}

 Thus, the inner behavior of the EYM black holes shows the following
 two unexpected features. The first is  `second quantization'
 of solutions with the `S' and `RN' type interiors:
 such solutions exist only for discrete values of the horizon radius.
 The second is an infinitely oscillating nature of the generic interior
 metric and an associated new type of singularity. No Cauchy horizon is
 formed in such a regime, although the blueshift at the local minima
 of  the mass function is extremely large and tends to infinity
 as singularity is approached. Mass function inflates exponentially during
 the first stage of each oscillation cycle and after some stabilization
 period comes back to nearly zero values. The upper bound of mass is
 infinitely growing at the singularity which is spacelike conformably
 with common expectations. However the singularity is not of the mixmaster
 type. It is described effectively by the second order dynamical system
 and hence is not chaotic, although huge metric oscillations are encountered.

 D. V. G. thanks the Theory Division, CERN for hospitality while
 the work was
 in progress. Stimulating discussions with G.~A.~Alekseev, I.~Bakas,
 G.~Cl\'ement,
 I.~G.~Dymnikova, P.~S.~Letelier, O.~I.~Mokhov, M.~S.~Volkov, and technical
 assistance of R.~N.~Zhukov are gratefully acknowledged.
 The research was supported in part by the RFBR grants 96--02--18899, 18126.


\begin{table}
\caption{S-- and RN--type solutions.}
\begin{tabular}{cccc}
 & S--type, $n=1$ & RN--type, $n=2$ & RN--type, $n=3$\\
\tableline
$r_h$     &$0.613861419$&$1.273791$&$1.0318420$\\
$W(r_h)$  &$-0.8478649145$&$-0.113763994$&$-0.10185163$\\
$r_{\_}$  & --- &$0.02171654$&$0.08948446$\\
$W(0)$    &$-1$&$-1.212296124$&$-1.3566052$\\
$\sigma(0)$&$0.2263801$&$5.991210\cdot10^{-3}$&$1.751928\cdot10^{-3}$\\
Mass      &$0.8807931$&$1.018002$&$1.000277$
\end{tabular}
\label{numres}
\end{table}

\begin{table}
\caption{Oscillations parameters for $r_h=2$, $W_h=-0.342072$
 ($n=1$ BH).}
\begin{tabular}{ccccc}
 &   $r$   & $W(r)$ & $W'(r)/r$ & $\Delta(r)$\\
\tableline
 $r^{max}_1$&$4.32\cdot10^{-2}$&$-0.881410$&27.89&$-2.52\cdot10^{-2}$\\
 $r^{min}_1$&$1.12\cdot10^{-2}$&$-0.922056$&67.03&$-1.68\cdot10^{-1}$\\
 $r^{max}_2$&$6.63\cdot10^{-4}$&$-0.926862$&$7.30\cdot10^3$&
 $-4.16\cdot10^{-4}$\\
 $r^{min}_2$&$4.73\cdot10^{-5}$&$-0.930103$&$1.49\cdot10^4$&
 $-8.12\cdot10^{36}$\\
 $r^{max}_3$&$6.44\cdot10^{-44}$&$-0.930120$&$4.10\cdot10^{81}$&
 $-1.36\cdot10^{-79}$\\
 $r^{min}_3$&$8.81\cdot10^{-83}$&$-0.930136$&$8.01\cdot10^{81}$&
 $-2\cdot10^{1.16\cdot10^{77}}$\\
 $r^{max}_4$&$2\cdot10^{-1.16\cdot10^{77}}$&$-0.930136$&
 $0.5\cdot10^{2.32\cdot10^{77}}$&$-2\cdot10^{-2.32\cdot10^{77}}$
\end{tabular}
\label{numosc}
\end{table}

\end{document}